\documentclass[RNAAS]{aastex63}

\graphicspath{{./}{figures/}}
\usepackage{graphicx}
\usepackage{float}

\begin{document}

\title{Modeling the quasiperiodic radial velocity variations of $\gamma$ Draconis}


\author{Victor Ramirez Delgado and Sarah Dodson-Robinson}
\affil{University of Delaware}


\begin{abstract}

$\gamma$~Draconis, a K5III star, showed radial velocity (RV) variations consistent with a 10.7 Jupiter mass planet from 2003-2011. After 2011, the periodic signal decayed, then reappeared with a phase shift.
\citet{hatzes18} suggested that $\gamma$~Dra's RV variations could come from oscillatory convective modes, but did not fit a mathematical model. Here we assess whether a quasi-periodic Gaussian process (GP)---appropriate when spots with finite lifetimes trace underlying periodicity---can explain the RVs. 
We find that a model with only one quasiperiodic signal is not adequate: we require a second component to fit the data.
The best-fit model has quasi-periodic oscillations with $P_1 = 705$~days and $P_2 = 15$~days. The 705-day signal may be caused by magnetic activity. 
The 15-day period requires further investigation.

\end{abstract}


\section{Project Goals} \label{sec:intro}

Doppler planet searches must contend with noise from magnetic activity, granulation, oscillations, and rotation, making planet candidate validation a challenging problem.
Here we reanalyze previously published radial velocity (RV) observations of $\gamma$~Dra from \citet{hatzes18}, which showed a planet-like signal that persisted for eight years before decaying and then re-emerging with a phase shift. We present a mathematical model of $\gamma$~Dra's RV variations and speculate on the physical processes underlying the apparent motion.

\section{RV Data and Jitter}
\label{sec:data}

The published RV data spans $>20$ years and include measurements from four different telescopes. To minimize zero-point offsets, we exclude data from the McDonald Observatory 2.1~m telescope from 1991--1993 and use only the data taken 
between 2003 and 2017.
Similar to \citet{hatzes18}, we find that the generalized Lomb-Scargle periodogram \citep{zechmeister09} has a peak at $682.6$ days, with false alarm probability $4.67 \times 10^{-65}$  \citep{baluev08} . The bottom panel of Figure \ref{fig1} shows the RV data plus randomly drawn realizations of the best-fit model (discussed in \S \ref{sec:optimize}).

Red giants have high granulation-induced RV jitter \citep{tayar19}. To find the RV error bars, we added the published errors and the expected jitter in quadrature. Expected jitter values come from 
\citet{yu18}\footnote{\texttt{https://github.com/Jieyu126/Jitter}} and are calculated for stellar parameters $\log{g} = 1.41 \pm 0.10$ (cgs), $T_{\rm eff} = 3965 \pm 28.0$~K \citep{2016A&A...587A...2B}, $L = 471 \pm 30 L_{\odot}$ \citep{2011A&A...526A.100P}, and $M = 1.72 \pm 1.02 M_{\odot}$ \citep{2011A&A...533A.107D}. 
For $\gamma$~Dra, the expected jitter is 19.86 m~s$^{-1}$.

\section{Model Framework} \label{sec:style}

Since the long-period signal is quasiperiodic (Figure \ref{fig1}, bottom), we model it with a Gaussian process, as in \citet{angus18}. 
Model 1 uses quasiperiodic kernel $k_1$:
\begin{equation}
k_1(i,j) = A \exp{\bigg[-\frac{(t_i - t_j)^2}{\lambda^2} -\Gamma \sin^2{\bigg(\frac{\pi (t_i - t_j)^2}{P}\bigg)}}\bigg] + \sigma_i^2\delta_{ij},
\label{first}
\end{equation}
where $k_1(i,j)$ is the covariance at times $(t_i, t_j)$, $A$ is the covariance amplitude, $P$ is the period, $\lambda$ is the decorrelation time of the periodic signal, and $\Gamma$ measures the smoothness of periodic variations. 
The term $\sigma_i^2 \delta_{ij}$ adds error bars to the diagonal of the covariance matrix.

When we found that Model 1 was unable to reproduce the observed RVs, we tested a second model. Model 2 retains the quasiperiodic kernel, but adds a second signal with period $P_2$ and roughness $\Gamma_2$:
\begin{equation}
k_2(i,j) = A \exp{\bigg[-\frac{(t_i - t_j)^2}{\lambda^2} -\Gamma_1\sin^2{\bigg(\frac{\pi (t_i - t_j)^2}{P_1}\bigg)} -\Gamma_2\sin^2{\bigg(\frac{\pi (t_i - t_j)^2}{P_2}\bigg)}}\bigg] + \sigma_i^2\delta_{ij}.
\label{second}
\end{equation}
As stars produce multiple (quasi)periodic signals that come from  combinations of activity cycles, rotation, and oscillations, it's no surprise that the covariance matrix required extra complexity to explain $\gamma$~Dra's RV variability.

\section{Parameter Optimization}
\label{sec:optimize}

\begin{figure}[ht]
\centering
\begin{tabular}{c}
\includegraphics[width=0.75\textwidth]{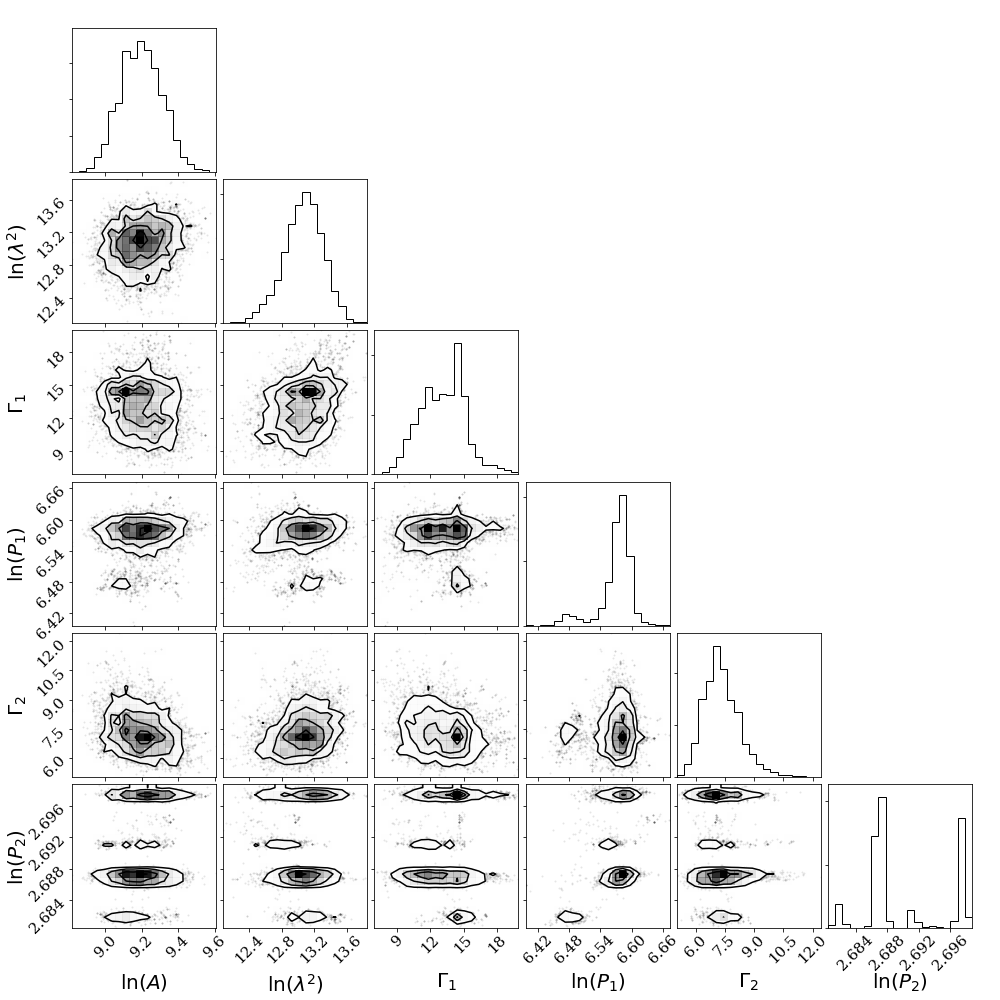} \\
\includegraphics[width=0.8\textwidth]{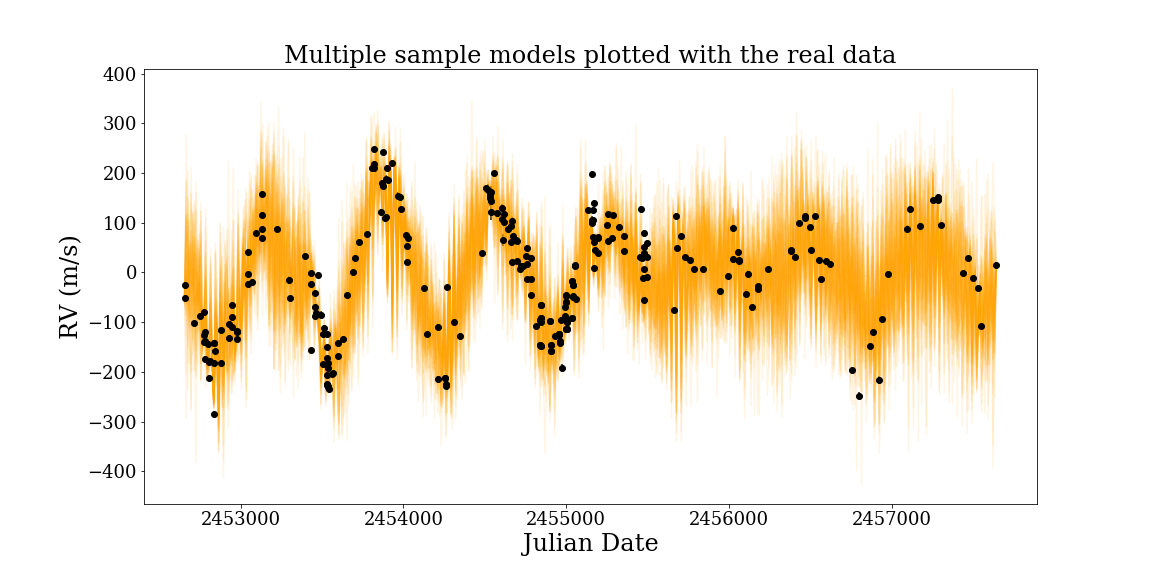} \\
\end{tabular}

\caption{{\bf Top}: Posterior probability distributions from the \texttt{emcee} sampler applied to Model 2 (Equation \ref{second}). 
The $P_2$ posterior distribution is bimodal, but the peaks are close enough together (14.68 and 14.82 days) to show that the RV variations have $P_2 \approx 15$~days.
{\bf Bottom:} \citet{hatzes18} $\gamma$~Dra RV data from 2003-2017 (black) plotted against 50 realizations of the best-fit GP model (orange).
}
\label{fig1}
\end{figure}

We used Bayesian inference to find the best-fit GP kernel parameters. We set up the Gaussian process with \texttt{George} \citep{2015ITPAM..38..252A} and used \texttt{emcee} \citep{2013PASP..125..306F} to perform a Markov-chain Monte Carlo (MCMC) exploration of parameter space.
When working with Model 1 (Equation \ref{first}), we found that the best-fit value of $\Gamma$ was always unreasonably high. With high $\Gamma$, data points that are separated in time by one period are highly correlated, but data points that are separated by {\it slightly} more or less than one period are almost uncorrelated. 
Figure \ref{fig1} shows that that the signal is coherent enough for the RV at time $t$ to have some predictive power on the RV at time $t + P_1 + \Delta t$, given $\Delta t \ll P_1$. We imposed stronger coherence by freezing $\Gamma$ in the \texttt{emcee} sampler. From a set of \texttt{emcee} simulations with $\Gamma = 1.0, 1.5, \ldots, 9.5, 10.0$, we found that the best-fit parameters are $P = 602.2$~days, $\lambda = 491.3$~days, $A = 96.54$~m~s$^{-1}$, and $\Gamma = 5.0$. Unlike Model 1, Model 2 was well-behaved, with the MCMC analysis yielding converged, reasonable parameter values. 
For Model 2, the best-fit parameters are $P_1 = 705.5$~days, $\lambda = 688.8$~days, $A = 98.25$~m~s$^{-1}$, $\Gamma_1 = 14.62$, $P_2 = 14.75$~days, and $\Gamma_2 = 7.18$.


To verify that the dataset had enough information to support choosing the more complex model (6 free parameters) over the simpler model (4 free parameters), we conducted two tests.
First, we evaluated the $\chi^2$ of 500 randomly-selected GP realizations from each model.
While the samples drawn from Model~1 yielded a reduced $\chi^2$ distribution centered at 2.65, Model~2 performed better, with a reduced $\chi^2$ distribution centered at 0.96.
Our second test was the Bayesian Information Criterion:
\begin{equation}\label{third}
    {\rm BIC} = m\ln{(n)} - 2\ln{(\mathcal{L})},
\end{equation}
where $m$ is the number of free parameters, $n$ is the number of observations, and $\mathcal{L}$ is the log-likelihood of a given parameter combination.
Again, we randomly chose 500 GP realizations from each model and examined the resulting BIC distributions.
For Model 1, the BIC distribution is centered at 3355; for Model 2, the distribution is centered at 3080. When two models yield $\Delta$BIC$ > 10$, the model with lower BIC is preferred. Here we have $\Delta$BIC$ = 275$. Model 2 is superior to Model 1.

\goodbreak
\section{Discussion}

Our last goal was to diagnose the cause of the RV variations.
As stated by \citet{hatzes18}, the long-period signal does not belong to a planet. What about the 15-day signal? Assuming a circular orbit, a planet with $P = 15$~days would be located $30.7 R_{\odot}$ from the center of the star. Using literature values of $\log{g}$ and $M$, we calculated a star radius of $R = 41.4 R_{\odot}$. 
There is no planet in a 15-day orbit, as such a planet would have been engulfed by the star. 

Another possibility is that one or both periodic signals come from rotation. Following \citet{surface}, we calculate the critical  period 
at which rotational kinetic energy overwhelms gravitational binding energy: $P\textsubscript{crit} = 43.2$~days. Clearly, $\gamma$~Dra cannot have a 15-day rotation period.
It's harder to assess whether rotation is causing the 705-day signal. \citet{surface} found that most of the {\it Kepler} red giants with detectable rotational modulation had periods of
50--150 days.
$\gamma$~Dra could be a slow rotator, 
but in that case it's 
hard to explain the high RV semiamplitude of 200~m~s$^{-1}$.


Lastly, we considered magnetic activity.
Aldebaran, a K5III giant like $\gamma$~Dra, has an RV signal with $P = 520$~days and amplitude $\sim 200$~m~s$^{-1}$, which \citet{alphatau} associate with magnetic activity based on simultaneous H$\alpha$ observations.
$\gamma$~Dra's 705-day period could come from magnetic activity, but we need to analyze activity indicators to confirm \citep{wise18}.

\newpage

\end{document}